\documentclass[aps,pra,twocolumn,final,letterpaper]{revtex4}

\usepackage{graphicx}   
\usepackage{import}                         
\usepackage{epstopdf}
\usepackage{amsmath} 
\usepackage{bm}
\usepackage{physics}
\usepackage{amssymb}
\usepackage{quotes}
\usepackage{color}
\usepackage{transparent}
\usepackage{dcolumn}

\usepackage{multirow}
\usepackage{cancel} 
\usepackage{mdframed}
\usepackage{color}
\usepackage{bm}
\usepackage{dsfont}
\usepackage{slashed}
\usepackage{soul, color} 
\soulregister\ref{7}  
\soulregister\cite{7} 
\renewcommand{\st}[1]{}

\usepackage{siunitx}
\usepackage{physics}

\DeclareMathOperator{\sinc}{sinc}
\begin{document}
\rmfamily

\title{Monochromatic X-ray source based on scattering from a magnetic nanoundulator}
\author{Sophie Fisher$^{1}$, Charles Roques-Carmes$^{2}$, Nicholas Rivera$^{1}$, Liang Jie Wong$^{3,4}$, Ido Kaminer$^{5}$, Marin Solja\v{c}i\'{c}$^{1}$}

\affiliation{$^{1}$ Department of Physics, Massachusetts Institute of Technology, Cambridge, MA 02139, USA. \\
$^{2}$ Research Laboratory of Electronics, Massachusetts Institute of Technology, Cambridge, MA 02139, USA. \\
$^{3}$ Singapore Institute of Manufacturing Technology, Singapore, Singapore. \\
$^{4}$ School of Electrical and Electronic Engineering, Nanyang Technological University, Singapore, Singapore. \\
$^{5}$ Department of Electrical Engineering, Technion, Haifa, Israel.
$\dagger$ Corresponding author e-mail: sefisher@mit.edu}

\noindent	

\begin{abstract}
We present a novel design for an ultra-compact, passive light source capable of generating ultraviolet and X-ray radiation, based on the interaction of free electrons with the magnetic near-field of a ferromagnet. Our design is motivated by recent advances in the fabrication of nanostructures, which allow the confinement of large magnetic fields at the surface of ferromagnetic nanogratings. Using \emph{ab initio} simulations and a complementary analytical theory, we show that highly directional, tunable, monochromatic radiation at high frequencies could be produced from relatively low-energy electrons within a tabletop design. The output frequency is tunable in the extreme ultraviolet to hard X-ray range via electron kinetic energies from 1 keV-5 MeV and nanograting periods from \SI{1}{\micro\metre}-5 nm. Our design reduces the scale, cost, and complexity of current free-electron-driven light schemes, bypassing the need for lengthy acceleration stages in conventional synchrotrons and free-electron lasers and driving lasers in other compact designs. Our design could help realize the next generation of tabletop or on-chip X-ray sources.

\end{abstract}

\maketitle

\noindent

Tabletop sources of extreme-ultraviolet and X-ray radiation are potentially useful for a wide variety of applications in medicine, engineering, and the natural sciences, ranging from medical therapy and diagnostics to X-ray imaging and spectroscopy, particle detection, and photolithography \cite{Sakdinawat2010, spectromicroscopy, doi:10.1063/1.4953071, Solak_2006}. Free-electron-driven light sources are promising schemes for realizing this goal. Synchrotrons and free-electron lasers are conventional sources that can generate high-quality, tunable radiation at extremely high brightnesses and intensities. Free-electron lasers are especially powerful due to their operation within a high-gain regime, in which the radiation intensity increases exponentially through coherent emission. These sources have enabled major advances in X-ray-based science, but their scope is severely limited by their large size and cost, driven by the need for kilometer-scale infrastructures to accelerate electrons to high energies, and complex undulator magnets with large field strengths to induce oscillating electrons \cite{McNeil2010, RevModPhys.88.015006}. Alternative compact, cheap sources using relatively low-energy electrons are required to realize the diverse spectrum of X-ray applications intended for laboratory and clinical settings. Recently, there have been a number of proposals for compact free-electron sources relying on various spontaneous emission effects, such as Cherenkov and transition radiation sources \cite{Liu2017} and Smith-Purcell emitters \cite{PhysRevLett.103.113901, Massuda:17, smith_purcell_low_energy, PhysRevX.7.011003, Roques-Carmes2019, Yang2018}, but these are limited to frequencies around the extreme ultraviolet where the material response becomes insubstantial. Graphene plasmon-based X-ray sources have also been proposed \cite{Wong2015, Rosolen2018}, but require a driving laser field and present technical challenges such as the intrinsically low interaction volume, as well as the need to sustain intense and highly-confined graphene-plasmon fields.

\begin{figure*}[htb]
\includegraphics[width=\textwidth]{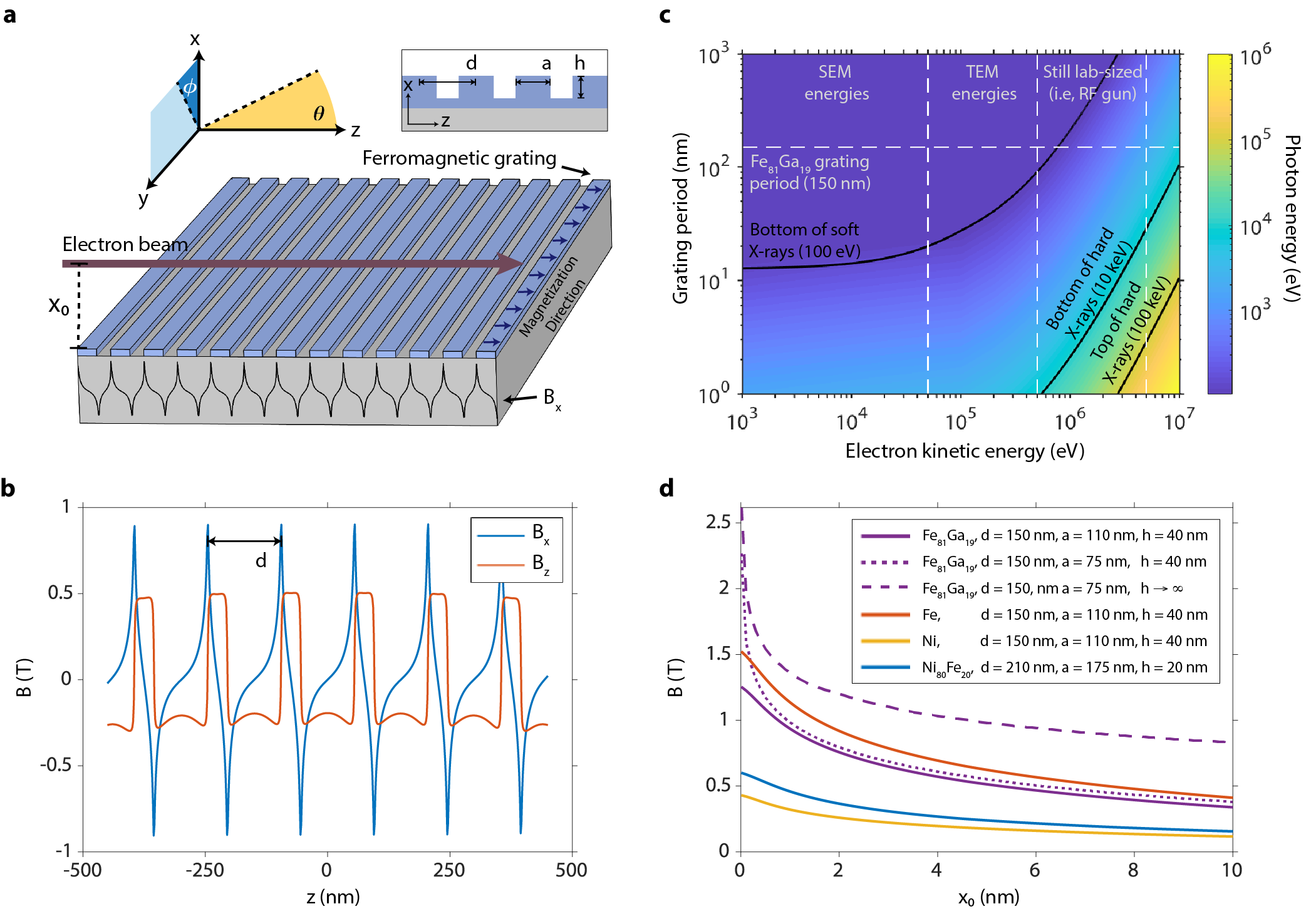}
\caption{\textbf{Compact free-electron source of high-frequency radiation. a},\ Schematic of the compact radiation source, in which free electrons oscillate in the magnetic near-field of a ferromagnetic nanograting to produce high-frequency radiation. The magnetic field is periodic in $z$ with the nanograting period and has high amplitude ($\sim 1-2$ T) towards the nanograting surface (small $x_0$) (\textbf{b}, \textbf{d}). \textbf{b}, Plots of the nonzero magnetic field components at a height $x_0 = 1$ nm above the surface, using the parameters of an $\textup{Fe}_{81}\textup{Ga}_{19}$ nanograting fabricated and charaterized in Ref.\ \cite{PhysRevB.97.060404}, where $\mu = 1.4 \times 10^6 \textup{A/m}$, $d = 150$ nm, $a = 110$ nm, and $h = 40$ nm. \textbf{d}, The parameters of the nanograting can be tuned to produce a stronger field at the surface by taking the filling ratio $a/d$ to 1/2 (purple, dotted line), increasing the height to period ratio $h/d$ (purple, dashed line), and choosing a ferromagnetic material with a higher magnetization at saturation (orange line). \textbf{c}, The dispersion relation of the source shows the production of high-frequency radiation ranging from extreme UV to hard X-rays that can be tuned with the nanograting period and electron kinetic energy. 
}

\label{one}
\end{figure*}

Here, we present a novel design for a compact, passive radiation source capable of generating ultraviolet and X-ray radiation, based on the interaction of free electrons with the magnetic near-field of a ferromagnetic material. Our design is motivated by recent advances in nanofabrication techniques of magnetic materials, which allow the confinement of large magnetic fields ($\sim$1 T) at the surface of nanopatterned ferromagnets \cite{PhysRevB.97.060404, PhysRevB.72.224413, 4544892, PhysRevB.73.235346, magnetic_nanostructures, Luttge_2009}. We show that the near-fields of ferromagnetic nanogratings can be leveraged to produce highly directional, tunable, monochromatic radiation at short wavelengths with relatively low energy electrons in a tabletop design. Our design offers a number of advantages over other free-electron sources, bypassing the need for lengthy acceleration stages in more conventional setups and driving laser fields in other compact designs. The output wavelength is tunable via the electron energy and nanograting period, with modest output powers that are suitable for small-scale applications. The output power could be enhanced by bunching the electron beam at the wavelength scale; for instance, an electron beam with $M = 10$ bunches is predicted to yield power in the range $1.3 \times 10^3$ to $1.1 \times 10^6$ photons s$^{-1}$ sr$^{-1}$ per $1 \%$ bandwidth for an electron energy range of 20 keV to 5 MeV. In the following, we describe the concept for our compact, high-frequency radiation source. We derive analytical expressions for the frequency and intensity of radiation due to the interaction of a single electron with the source. Our expressions yield excellent predictions of the radiation output, as confirmed by our \emph{ab initio} numerical simulations. We also calculate the radiation due to free-electron beams from conventional, laboratory-sized sources and discuss how the output power can be maximized by tuning the nanograting and electron beam geometries. Finally, we show how the power can be enhanced up to a factor of $M^2$ by bunching the electron beam at the wavelength scale, where $M$ is the number of bunches. 

The mechanism behind the electromagnetic radiation source is illustrated in Fig.\ \ref{one}a. The setup contains a grating with nanoscale periodicity fabricated from a ferromagnetic material. The nanograting is prepared to have a magnetization along the $z$ direction, transverse to the nanograting stripes, by exposing the structure to an external homogeneous magnetic field. When the external magnetic field is turned off, the ferromagnetic material exhibits its saturation magnetic field due to magnetic hysteresis. The resulting field is periodic along $z$ with the nanograting period and falls off with increasing height ($x_0$) above the nanograting (Fig.\ \ref{one}d). Electrons are sent parallel to the nanograting in the direction of magnetization. Interaction with the magnetic field causes electrons to oscillate primarily in the directions transverse to its unperturbed motion, which leads to the generation of high-frequency, directional radiation. The mechanism can be thought of as inverse-Compton scattering, whereby in the rest frame of the electron, a virtual photon from the magnetic field scatters off of the electron and gains energy. The same principle on a larger scale applies to undulator radiation from free-electron lasers, where the undulator period is typically a few centimeters, the magnetic field strength is 1 T, and the electron energy is 1-15 GeV. In our case, the nanoundulator period reduces to around 100 nanometers and the electron energy to 1 keV-5 MeV, with the field strength remaining at 1 T. Since the fundamental wavelength of radiation is proportional to $d/ \gamma^2$, where $d$ is the undulator period and $\gamma$ is the relativistic Lorentz factor for the electrons, our design reaches comparable frequencies to free-electron lasers with much lower initial electron energies. 

An analytic formula for the magnetic field $ \vb{B}(x,y,z)$ created by a nanoundulator has been previously derived \cite{PhysRevB.73.235346}. The expression is computed by an integration of the magnetic field produced by a magnetic dipole moment over the spatial distribution of the nanograting, assuming a constant magnetization $\mu$ in the $z$ direction. The nanograting is taken to be infinite in the $y$ direction and finite in the $z$ direction. The resulting magnetic field has $B_{y}(x,y,z) = 0$, with $B_{x}(x,y,z)$ and $B_{z}(x,y,z)$ periodic in $z$ with the nanograting period for $x$ close to the nanograting surface (small $x_0$). The exact profiles for $B_{x}$ and $B_{z}$  have dependences on four parameters of the nanograting: $d$, the nanograting period, $a$, the stripe width, $h$, the stripe height, and $\mu$, the magnitude of magnetization at saturation, a function of the ferromagnetic material. In particular, the field components depend upon the ratios of the geometric parameters, and $\mu$ is a scale factor on each component. Throughout the article, all calculations are carried out for a ferromagnetic nanograting composed of Galfenol $(\textup{Fe}_{81}\textup{Ga}_{19})$, with $\mu = 1.4 \times 10^6 \textup{A/m}$, $d = 150$ nm, $a = 110$ nm, and $h = 40$ nm, unless stated otherwise. Such a structure was previously fabricated for a different study through magnetron sputtering and focused ion beam milling \cite{PhysRevB.97.060404}. The choice of our reference nanograting is motivated by its relatively high magnetization, which induces a strong magnetic field, and the nanoscale period, which facilitates the production of high frequency radiation.

Fig.\ \ref{one}b plots the components of the magnetic field induced by the reference nanograting as a function of $z$ at a height of $x_0 = 1$ nm. $B_x$ and $B_z$ are periodic along $z$ with period $d$, and both components peak near fields of 1 T. Such an amplitude corresponds to the same amplitude of an electric field of intensity 0.3 TW cm$^{-2}$, meaning that a laser undulator would require this level of intensity to provide a similar power per electron. As the out-of-plane distance to the nanograting ($x_0$) increases, the magnetic field magnitude exhibits rapid decay, as illustrated in Fig.\ \ref{one}d. Fig.\ \ref{one}d also shows how the field magnitude can be tuned via the nanograting parameters, with plots of the field of the reference nanograting (solid purple line) against that of other structures. For instance, when the stripe width of the reference nanograting is changed from $a = 110$ nm to $a = 75$ nm, the magnitude is increased by $\sim$1 T at the surface of the nanograting (dotted purple line). At this width, the field is optimized with respect to $a$ for fixed $d$, and the filling ratio $a/d$ is precisely 1/2. When the height of the reference nanograting is taken from $h = 40$ nm to the limit where $h \rightarrow \infty$, the field is optimized with respect to $h$ and the magnitude is further increased (dashed purple line). In a third variation, the material of the reference nanograting is changed to Fe (orange line), which results in a slightly larger field due to the higher magnetization of Fe, $\mu = 1.7 \times 10^6$ A/m. For reference, we include plots for two more nanogratings composed of Ni (yellow line) and $\textup{Ni}_{80}\textup{Fe}_{20}$ (blue line), which have been previously fabricated for other experiments using deep ultraviolet lithography \cite{PhysRevB.72.224413, 4544892}.

\begin{figure*}[htb]
\includegraphics[width=\textwidth]{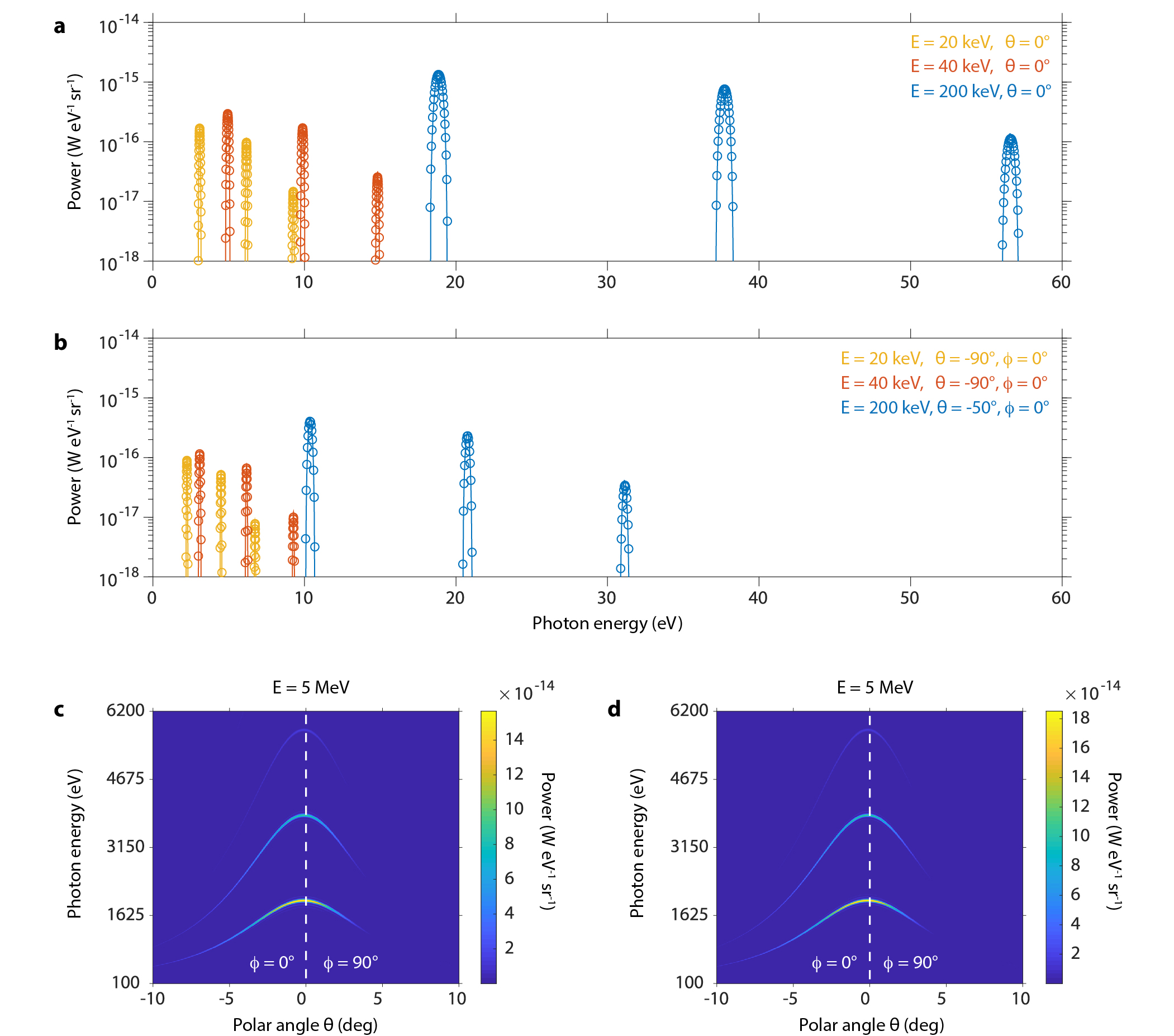}
\caption{\textbf{Tunability of the emitted photon energy via the electron energy. a}, \textbf{b},\ Numerical (circles) vs.\ analytical (solid lines) results of the radiation power from a single electron in units of power per photon energy per solid angle for a variety of electron energies and fixed polar angles. Different colors represent different electron energies. The polar angles in \textbf{b} are chosen such that $\phi = 0$ and $\theta$ is approximately halfway between the leftmost edge and the peak of the full angular spectrum. \textbf{c}, \textbf{d},\ Numerical (\textbf{c}) vs analytical (\textbf{d}) results of the full angular spectrum of radiation power from a single electron with energy 5 MeV. All calculations are carried out using the parameters of the reference nanograting  \cite{PhysRevB.97.060404} with a length along $z$ of \SI{4.5}{\micro\metre} and an electron launched in the $z$ direction with initial height $x_0 = 1$ nm above the nanograting surface.} 
\label{two}
\end{figure*}

We now turn to developing an analytical theory which describes the output radiation due to a single electron interacting with the magnetic near-field of the ferromagnetic nanograting. Our theory yields an excellent prediction of the frequency and intensity of output radiation, as confirmed by comparisons to our \emph{ab initio} numerical simulations (see Fig.\ \ref{two}). The dispersion relation of the system can be derived using the law of energy-momentum conservation. We consider an elastic collision between an electron of rest mass $m$ launched in the $z$ direction with velocity $v$ (normalized velocity $\beta = v/c, \textup{Lorentz factor} \ \gamma = (1 - \beta^2)^{-1/2})$ and a nanograting yielding a momentum $2\pi \hbar n / d$, where $d$ is the nanograting period. The output photon has departing angle $\theta$ with energy $\hbar \omega$ and momentum $\hbar \omega / c$. Then $\omega$ is given by  \begin{equation} 
\omega \approx \frac{  2 \pi c n}{d(1 - \beta \cos{\theta})} \label{dispersion}. 
\end{equation} 
The approximate equality neglects the effects of quantum recoil and holds whenever $\frac {2\pi \hbar n}{d} \ll \gamma \beta m c$, which is always true in the case under study.

We now derive a fully analytical expression that predicts the spectral power of emitted radiation as a function of its frequency $\omega$, azimuthal angle $\phi$, and polar angle $\theta$. We assume the electron is launched sufficiently close to the surface of the nanograting such that the fields are periodic with the nanograting period $d$. We also assume that the electron's transverse velocity oscillations are small enough such that $\gamma$ and $\beta$ are approximately constant throughout the interaction, i.e. $\dot{z} \approx v$ and $z \approx vt + z_0$. In this case, the Lorentz forces on the electron due to $B_z$ are small, and $B_z$ can be disregarded. We then write the magnetic field as $\vb{B}(x,y,z) = \sum\limits_{n} B_n \sin{(\frac{2  \pi  z}{d_n})}\hat{x}$ where we have Fourier expanded $B_x$, and where $B_n$ is the amplitude and $d_n = d/n$ is the period of the $n$-th Fourier component. Only the dominant Fourier components contribute to the radiation spectrum, with each contributing to spectrally distinct peaks (see Fig.\ \ref{two}). We find
\begin{equation}
\frac{\dd^2P}{\dd \Omega \dd \omega} = \frac{e^2 \omega^2 T }{64 \pi^3 c \varepsilon_0} \big| U \big|^2,
\label{final} \end{equation}
where
\begin{equation}
U =  \sum\limits_{n=1}^{\ell} \big( \vb{B} K_n - \vb{A} \xi_n \big) \sinc{\bigg( \frac{ T [\omega(1 - \beta \cos{\theta}) - W_n] }{2}\bigg)},
\end{equation}
\begin{equation} \vb{A} = 
	\begin{pmatrix} 
	\beta \sin{\theta} \cos{\theta} \cos{\phi} \\
	\beta \sin{\theta} \sin{\phi} \cos{\theta} \\
	-\beta \sin^2{\theta}
	\end{pmatrix},
	\vb{B} = \begin{pmatrix}
	\sin^2{\theta} \sin{\phi} \cos{\phi} \\
	-\sin^2{\theta}\cos^2{\phi} - \cos^2{\theta}  \\
	\sin{\theta} \sin{\phi} \cos{\theta}
	\end{pmatrix},
\end{equation}
where $\xi_n = \frac{ -K_n \omega \sin{\theta} \sin{\phi} }{W_n \gamma}$ is the amplitude of oscillation in the $y$ direction normalized to the $y$ component of the wave vector $k_y = \frac{\omega \sin{\theta} \sin{\phi}}{c}$, $K_n = \frac{e B_n d_n}{2 \pi m c} $ is the magnetic deflection parameter and $W_n = \frac{2 \pi \beta c}{d_n}$ is the frequency of oscillation due to the $n$-th Fourier component, $e$ is the electron charge, $\varepsilon_0$ is the permittivity of free space, $T = \frac{Nd}{v}$ is the electron flight time, and $N$ is the number of nanograting periods. The full derivation is provided in Supplementary Section I. Exact numerical simulations of the interaction for a range of initial electron energies were carried out and directly compared to the analytics. The results show excellent agreement and confirm the validity of our analytical theory (see Fig.\ \ref{two}).

Fig.\ \ref{one}c plots the dispersion relation from Eq.\ \ref{dispersion} (for $n=1$), showing how the output radiation frequency can be tuned with the electron kinetic energy and nanograting period. Contour lines in black denote the boundaries of hard and soft X-rays, and white vertical lines separate various regimes of electron energies. The white horizontal line marks the nanograting period corresponding to the reference geometry (150 nm). The two leftmost regimes of electron energies correspond to nonrelativistic electrons from 1-500 keV that can be realized in conventional scanning electron microscopes (SEM) and transmission electron microscopes (TEM). These regimes are already sufficient for hard ultraviolet generation with 150 nm nanogratings, and soft X-rays can be reached with smaller periods, with $20$ nm nanogratings at 50 keV, and $90$ nm nanogratings at $500$ keV. The next regime corresponds to electron energies from 500 keV-5 MeV, achievable from conventional radio frequency (RF) guns (still lab-sized). At 1 MeV, soft X-rays are already achievable with 150 nm nanogratings. Hard X-rays can be reached with 11 nm nanogratings at 3 MeV and 29 nm nanogratings at 5 MeV. Current nanofabrication techniques such as electron-beam lithography can already pattern ferromagnetic materials with features as small as 10 nm \cite{magnetic_nanostructures}, and sub-5 nm features have been demonstrated for metallic materials \cite{Manfrinato2017}. With a 10 nm period, our design reaches soft and hard X-rays within a laboratory setup, circumventing the need for high energy particle accelerators and kilometer-scale infrastructures, and vastly reducing the scale and cost of most free-electron X-ray sources. 

\begin{figure*}[htb]
\includegraphics{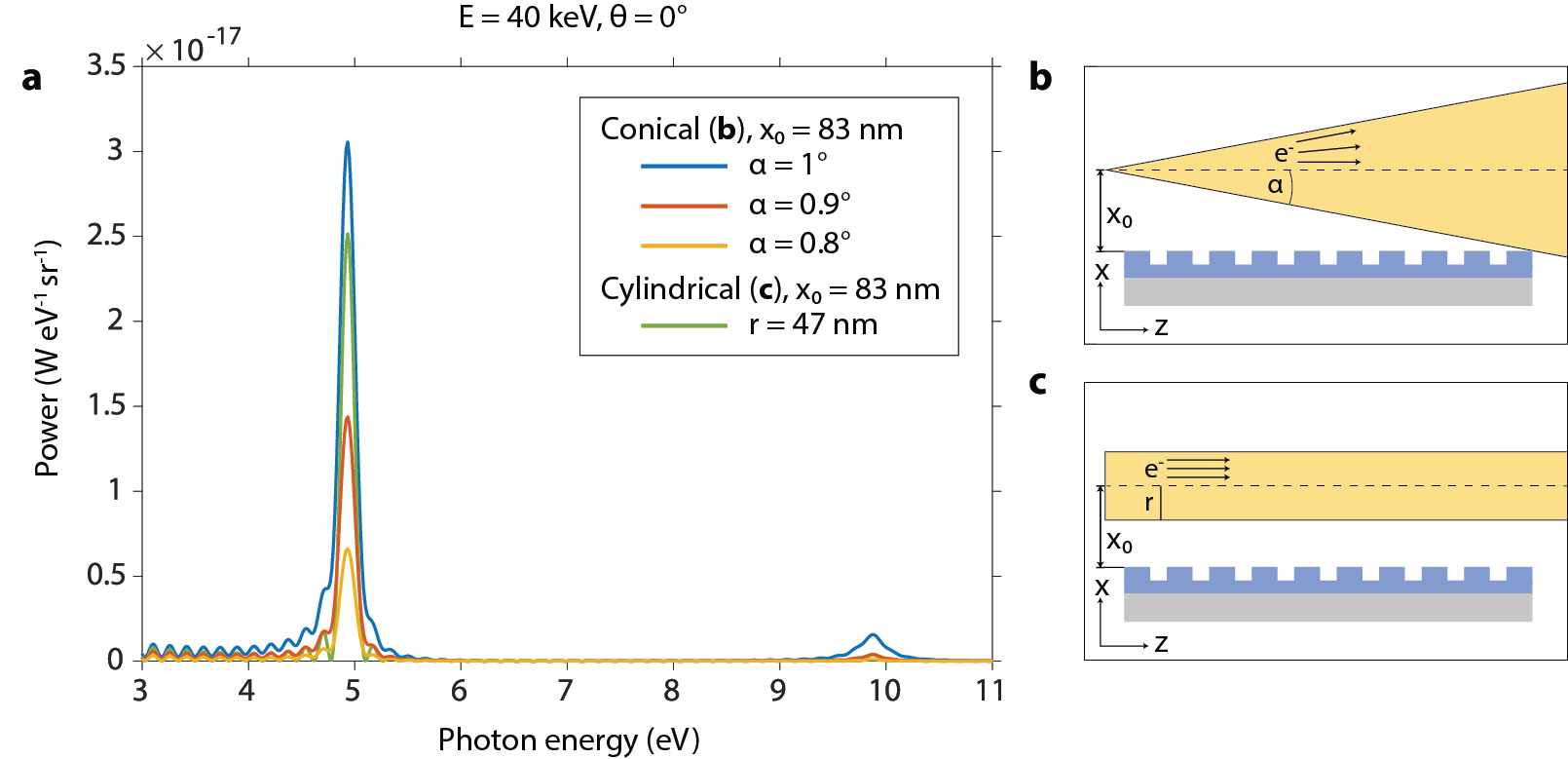}
\caption{\textbf{Dependence of the power spectrum on electron beam geometry. a}.\ Numerical results of the on-axis radiation power for conical (\textbf{b}) and cylindrical (\textbf{c}) electron beam geometries in units of power per photon energy per solid angle. Conical beams with varying half-angles $\alpha$ are considered. The initial height $x_0$ for all geometries is chosen so that the conical beam with the largest half-angle ($\alpha = 1^{\circ}$) grazes the edge of the nanograting, as in \textbf{b}. The electron energy is set to 40 keV and the beam brightness to $10^9$ A cm$^{-2}$ sr$^{-1}$. All nanograting parameters are as in Fig.\ \ref{two}.} 
\label{three}
\end{figure*}

Next, we discuss the results of our \emph{ab initio} numerical simulations for single electrons of varying energies interacting with the nanograting. The simulations compute exact radiation spectra due to the interaction by solving for electron trajectories using the Newton-Lorentz equation and taking Fourier transforms of the fields obtained from the Li\'enard-Wiechert potentials \cite{Wong2015, Wong2017}. We perform simulations over a range of polar angles and initial electron energies, taking the parameters of the reference nanograting. We set the length of the nanograting along $z$ to be $Nd = \SI{4.5}{\micro\metre}$, where $N = 31$. The results are displayed in Fig.\ \ref{two} (a \& b solid lines), along with a comparison to our analytical formula (Eq.\ \ref{final}, a \& b circles), and show the production of highly directional, monoenergetic radiation. Each spectrum displays the first three harmonics of radiation generated from the dominant Fourier components of the magnetic field. There is a clear dependence of output power density on electron energy, with the scaling factor of $\gamma^2$ evident from the analytics (Eq.\ \ref{dispersion} \textup{and} \ref{final}), as $\omega \sim \gamma^2$ for on-axis peaks ($\theta = 0^{\circ}$) when $\beta \approx 1$, and $\dd \omega \sim \gamma^2$. In Fig.\ \ref{two}a, we plot the on-axis power per photon energy per solid angle (W eV$^{-1}$ sr$^{-1}$) for three different electron energies. Electrons with a kinetic energy of 40 keV, which are achievable with some SEM models, generate ultraviolet photons with on-axis peak energies of 4.96 eV ($2.63 \%$ FWHM energy spread), 9.89 eV ($1.32 \%$ spread), and 14.83 eV ($0.81 \%$ spread). At a kinetic energy of 200 keV, achievable with a TEM, electrons produce extreme ultraviolet photons with on-axis peak energies of 18.86 eV (2.6$\%$ spread), 37.71 eV ($1.19 \%$ spread), and 56.55 ($0.80 \%$ spread). In Fig.\ \ref{two}b, we display the same plot with different fixed polar angles $\theta$ and $\phi$, showing the angular dependence of the output photon energy. We choose $\theta$ to be approximately halfway between the leftmost edge and the peak of the full angular spectrum (see Fig.\ \ref{two}c and \ref{two}d). Fig.\ \ref{two}c shows the numerical result of the full angular spectrum for an electron kinetic energy of 5 MeV, achievable with a lab-sized RF gun. The spectrum shows the generation of soft and hard X-ray photons, with on-axis peak energies of 1.91 keV (2.56 $\%$ spread), 3.82 keV (1.17 $\%$ spread), and 5.74 keV (0.85 $\%$ spread). The radiated photons have an angular spread of $\sim 20^{\circ}$ since $\Delta \theta \sim 1/\gamma$. In Fig.\ \ref{two}d, we plot the analytical result for the angular spectrum from a 5 MeV electron, which shows very good agreement with \ref{two}c.

\begin{figure*}[htb]
\includegraphics{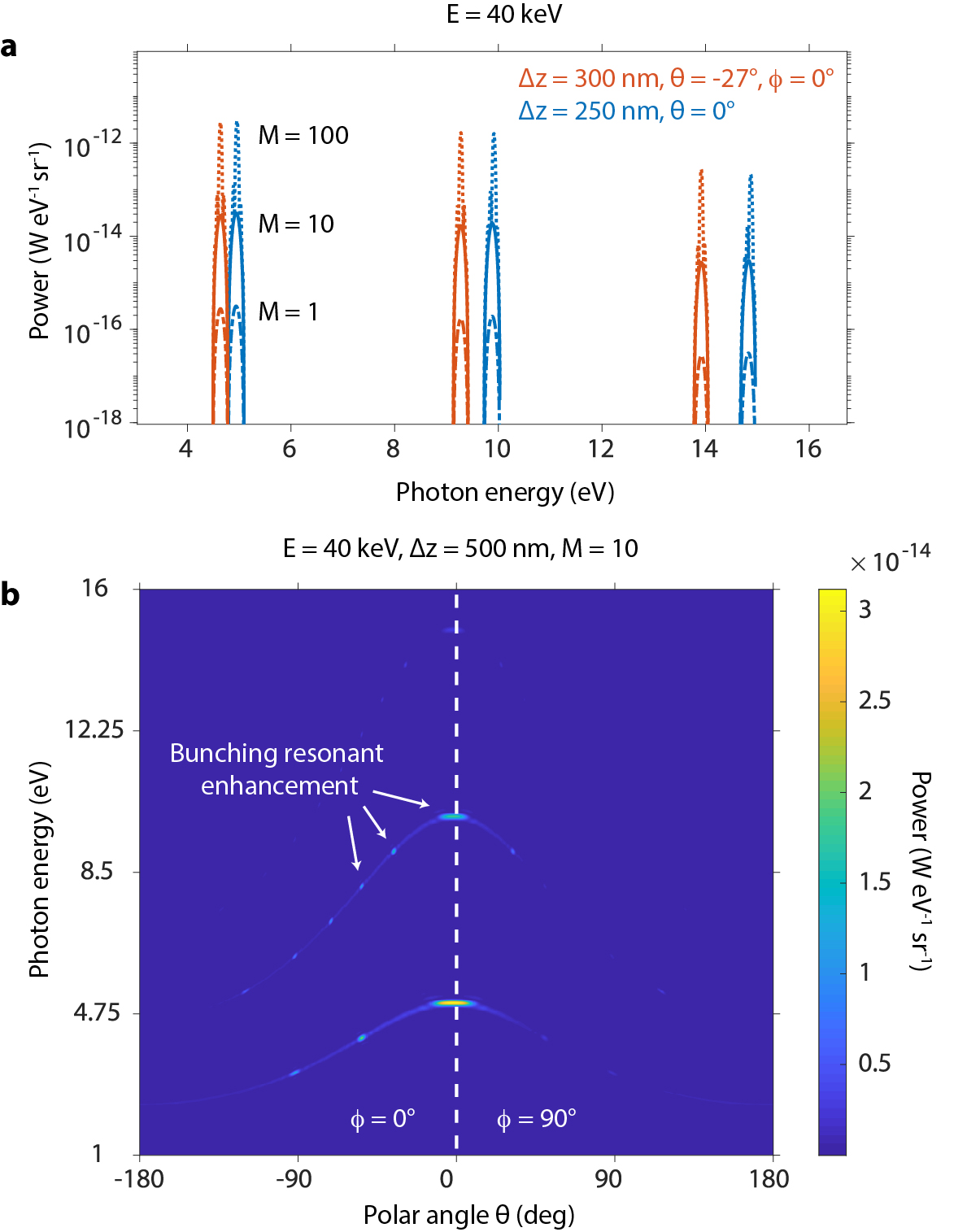}
\caption{\textbf{Enhanced power spectrum due to a bunched electron beam. a}, \textbf{b}, Radiation power in units of power per photon energy per solid angle due to a bunched electron beam with uniform energy 40 keV launched in the $z$ direction with initial height $x_0 = 1$ nm above the nanograting. The beam is one electron thick and consists of $M$ electrons linearly spaced along $z$ by a distance $\Delta z$. \textbf{a},\ The radiation power for varying values of $\Delta z, M$, and polar angle $\theta$. The polar angles are chosen so that $\phi = 0^{\circ}$ and $\theta$ lies on the enhancement occuring closest to $\theta =  0^{\circ}$. The total radiation power increases by two orders of magnitude for every order of magnitude increase in $M$, the number of bunches. \textbf{b},\ The full angular spectrum of radiation power for a 40 keV electron beam with $\Delta z =$ 250 nm and $M = 10$, showing enhancement both on and off axis. }
\label{four}
\end{figure*}

We now discuss the radiation due to free-electron beams from conventional, laboratory-sized sources and show how the output power can be enhanced by tuning the nanograting and electron beam geometries. First, it is emphasized that choosing a nanograting with optimal parameters for generating a large magnetic near-field is essential, as the power scales with the field strength (Eq.\ \ref{final}). These include a ferromagnetic material with a large magnetization, a stripe width to period ratio $a/d$ of $1/2$, and a high stripe height to period ratio $h/d$ (see Fig.\ \ref{one}d). Other methods of enhancement include increasing the length of the nanograting along $z$, since the power scales with the electron interaction time $T$ (Eq.\ \ref{final}), and vertically stacking multiple nanogratings in parallel. 

Lab-sized electron guns typically produce electron beams with a conical profile, with an outgoing half-angle $\alpha$ (Fig.\ \ref{three}b). Larger values of $\alpha$ yield higher beam currents but result in less power radiated per electron. In such a setup, the value of $\alpha$ should be chosen at around a few degrees to balance this tradeoff. Given a fixed value of $\alpha$, the initial height of the beam $x_0$ should be chosen so that electrons graze the edge of the nanograting (see Fig.\ \ref{three}b) to minimize distance to the surface and prevent surface collisions. Fig.\ \ref{three}a plots the on-axis power per photon energy per solid angle for a few conical beams with differing $\alpha$, obtained via numerical simulation. The electron energy is set to 40 keV and the beam brightness to $10^9$ A cm$^{-2}$ sr$^{-1}$, parameters obtainable in a cold field emission SEM \cite{wet-STEM}. The initial height $x_0 = 83$ nm for all beams is chosen so that the beam with the largest half-angle considered ($\alpha = 1^{\circ}$) grazes the edge of the nanograting. Fig.\ \ref{three}a shows that the conical beam with $\alpha = 1^{\circ}$ maximizes the on-axis output power, yielding 2.3 photons s$^{-1}$ sr$^{-1}$ per $1 \%$ bandwidth for a photon energy of 4.96 eV, and the power drops off with decreasing $\alpha$ (red and yellow lines). Fig.\ \ref{three}a also plots the on-axis power distribution for a cylindrical beam with initial height $x_0 = 83$ nm (green line, Fig.\ \ref{three}c), yielding similar values to the $\alpha = 1^{\circ}$ cylindrical case. For reference, when the electron energy is 5 MeV, the $\alpha = 1^{\circ}$ conical geometry yields an on-axis output power of $3.8 \times 10^2$ photons s$^{-1}$ sr$^{-1}$ per $1 \%$ bandwidth for a photon energy of 1.3 keV.

One of the novel features of our source is its ability to produce tunable, directional, monochromatic radiation at comparable frequencies to conventional free-electron sources with drastically lower electron energies. One downside to using low-energy electrons is a decrease in the output power, since the power density scales with $\gamma^2$ (Eq.\ \ref{dispersion} \textup{and} \ref{final}). The electrons used with our source typically have $\gamma \approx$ 1-10 and cannot match the output power of larger, more conventional X-ray sources, where $\gamma \gg 1$. Additionally, low-energy ($< 5$ keV) electron beams typically achieved in SEMs exhibit large angular divergences, thus reducing the efficiency of the radiation process \cite{xray}. Nevertheless, we proceed to point out that it is possible to enhance the output intensity substantially through the use of flat electron beams and bunched electrons.

Since the magnetic field of the nanograting drops off rapidly above the surface, using high-quality, high-brightness electron beams focused as close as possible to the source will increase the photon yield. Beams that are flattened along the $y$-$z$ plane allow a greater number of electrons to lie closer to the nanograting surface and experience a stronger field \cite{PhysRevSTAB.4.053501, PhysRevSTAB.9.031001}. Further, the output power can be enhanced orders of magnitude by pre-bunching the electron beam at the wavelength scale of radiation to produce coherent emission. In Supplementary Section II, we derive the spectral power of emitted radiation due to a bunched electron beam interacting with the source, using Ref.\ \cite{PhysRevSTAB.11.061301}. We consider a beam which is one electron thick, consisting of $M$ equally spaced electrons along $z$ by a distance $\Delta z$. The beam is launched parallel to the nanograting in the $z$ direction with uniform initial velocity $v$. Then, the output power of radiation is given by 
\begin{equation}
\frac{\dd^2P_{\text{tot}}}{\dd \Omega \dd \omega} = M[1 + (M-1) \big|f\big|^2] \frac{\dd^2P_{\text{sing}}}{\dd \Omega \dd \omega},
\label{bunching}
\end{equation}
where
\begin{equation}
\big|f\big|^2 = \frac{1}{M^2} \frac{\sin^2 \big( { \frac{M \omega \cos{(\theta)} \Delta z }{2 c}  } \big) }{ \sin^2 \big({ \frac{ \omega \cos{(\theta)} \Delta z }{2 c} }\big) },
\label{bunching2}
\end{equation}
where $f$ is the Fourier transform of the charge distribution and $\frac{\dd^2P_{\text{sing}}}{\dd \Omega \dd \omega}$ is the power due to a single electron, given by Eq.\ \ref{final}. The power enhancement is strongest where the bunching resonant enhancement condition $\big|f\big|^2 = 1$ holds, or where $\omega \cos{\theta} = \frac{2 c m \pi}{\Delta z}$ for integers $m$. When this holds, $\frac{\dd^2P_{\text{sing}}}{\dd \Omega \dd \omega} = M^2 \frac{\dd^2P_{\text{sing}}}{\dd \Omega \dd \omega}$. In Fig.\ \ref{four}, we plot the enhanced power spectrum due to Eq.\ \ref{bunching} for a fixed uniform beam energy of 40 keV. Fig.\ \ref{four}a shows line cuts of the spectrum for a few values of $\Delta z$ and $M$. On-axis peaks are enhanced when $\Delta z = 250$ nm, and peaks slightly off-axis ($\theta = -27^{\circ}, \phi = 0^{\circ}$) are enhanced when $\Delta z = 300$. The radiation power increases by two orders of magnitude when $M$ goes from 1 to 10, and another two orders of magnitude when $M$ goes to 100. For instance, when $\Delta z = 250$ nm, the power per solid angle of the on-axis peak at 4.96 eV goes from $1.7 \times 10^1$ photons s$^{-1}$ sr$^{-1}$ per $1 \%$ bandwidth to $1.6 \times 10^3$ photons s$^{-1}$ sr$^{-1}$ per $1 \%$ bandwidth when $M = 10$, and to  $1.2 \times 10^5$ photons s$^{-1}$ sr$^{-1}$ per $1 \%$ bandwidth when $M = 100$. In Fig.\ \ref{four}b, we show the full angular spectrum due to a bunched electron beam for $\Delta z = 500$ nm and $M = 10$. The enhancement is strongest on-axis with an angular spread of $\sim 24^{\circ}$, but occurs at off-axis angles where the bunching resonant enhancement condition holds. Additional off-axis angles can be enhanced (while keeping the on-axis peaks enhanced) by changing $\Delta z$ to a higher multiple of 250 nm, as evident from the bunching resonant enhancement condition.

In conclusion, we have presented the concept for a compact, passive light source that produces highly directional, monochromatic radiation. The light is tunable via the ferromagnetic nanograting period and the electron energy, spanning the extreme ultraviolet to hard X-ray range. The output power can be enhanced through a number of methods, such as shaping the electron beam, bunching the beam at the wavelength scale, increasing the length of the nanograting, and stacking multiple nanogratings in parallel. The source is well suited to laboratory-scale applications with its small size, low complexity, and low cost, requiring neither large high-energy accelerators nor driving laser fields. Ferromagnetic nanogratings with feature sizes as small as 10 nm can be produced with lithographic methods. Future advances in nanofabrication could allow the patterning and mass production of ferromagnetic nanogratings at sub 5-nm scales, making even higher photon energies achievable. Further, the small scale of the source makes it suited to an on-chip device. Recent progress has been made in high-energy on-chip accelerators \cite{RevModPhys.86.1337, Peralta2013}, which, when combined with our design, could enable an ultra-compact, fully on-chip X-ray light source. Finally, an interesting topic for future research is the possibility of a nanoscale free-electron laser, in which electrons bunch coherently via self-amplified spontaneous emission, allowing for an exponential increase in radiation power.
\\
\begin{acknowledgments} Nicholas Rivera acknowledges the support of the DOE Computational Science Graduate Fellowship (CSGF) Number DEFG02-97ER25308. L. J.W. was supported by the Advanced Manufacturing and Engineering Young Individual Research Grant (No. A1984c0043) from the Science and Engineering Research Council of the Agency for Science, Technology and Research, Singapore. This material is based upon work supported in part by the U.S. Army Research Office through the Institute for Soldier Nanotechnologies, under contract number W911NF-18–2–0048.
\end{acknowledgments}

\bibliography{magnetic_compton_bib.bib}

\begin{thebibliography}{30}
\expandafter\ifx\csname natexlab\endcsname\relax\def\natexlab#1{#1}\fi
\expandafter\ifx\csname bibnamefont\endcsname\relax
  \def\bibnamefont#1{#1}\fi
\expandafter\ifx\csname bibfnamefont\endcsname\relax
  \def\bibfnamefont#1{#1}\fi
\expandafter\ifx\csname citenamefont\endcsname\relax
  \def\citenamefont#1{#1}\fi
\expandafter\ifx\csname url\endcsname\relax
  \def\url#1{\texttt{#1}}\fi
\expandafter\ifx\csname urlprefix\endcsname\relax\def\urlprefix{URL }\fi
\providecommand{\bibinfo}[2]{#2}
\providecommand{\eprint}[2][]{\url{#2}}

\bibitem[{\citenamefont{Sakdinawat and Attwood}(2010)}]{Sakdinawat2010}
\bibinfo{author}{\bibfnamefont{A.}~\bibnamefont{Sakdinawat}} \bibnamefont{and}
  \bibinfo{author}{\bibfnamefont{D.}~\bibnamefont{Attwood}},
  \bibinfo{journal}{Nature Photonics} \textbf{\bibinfo{volume}{4}},
  \bibinfo{pages}{840} (\bibinfo{year}{2010}),
  \urlprefix\url{https://doi.org/10.1038/nphoton.2010.267}.

\bibitem[{\citenamefont{Hitchcock}(2015)}]{spectromicroscopy}
\bibinfo{author}{\bibfnamefont{A.~P.} \bibnamefont{Hitchcock}},
  \bibinfo{journal}{Journal of Electron Spectroscopy and Related Phenomena}
  \textbf{\bibinfo{volume}{20}}, \bibinfo{pages}{49} (\bibinfo{year}{2015}).

\bibitem[{\citenamefont{Schmitz et~al.}(2016)\citenamefont{Schmitz, Wilson,
  Rudolf, Wiemann, Plucinski, Riess, Schuck, Hardtdegen, Schneider, Tautz
  et~al.}}]{doi:10.1063/1.4953071}
\bibinfo{author}{\bibfnamefont{C.}~\bibnamefont{Schmitz}},
  \bibinfo{author}{\bibfnamefont{D.}~\bibnamefont{Wilson}},
  \bibinfo{author}{\bibfnamefont{D.}~\bibnamefont{Rudolf}},
  \bibinfo{author}{\bibfnamefont{C.}~\bibnamefont{Wiemann}},
  \bibinfo{author}{\bibfnamefont{L.}~\bibnamefont{Plucinski}},
  \bibinfo{author}{\bibfnamefont{S.}~\bibnamefont{Riess}},
  \bibinfo{author}{\bibfnamefont{M.}~\bibnamefont{Schuck}},
  \bibinfo{author}{\bibfnamefont{H.}~\bibnamefont{Hardtdegen}},
  \bibinfo{author}{\bibfnamefont{C.~M.} \bibnamefont{Schneider}},
  \bibinfo{author}{\bibfnamefont{F.~S.} \bibnamefont{Tautz}},
  \bibnamefont{et~al.}, \bibinfo{journal}{Applied Physics Letters}
  \textbf{\bibinfo{volume}{108}}, \bibinfo{pages}{234101}
  (\bibinfo{year}{2016}), \eprint{https://doi.org/10.1063/1.4953071},
  \urlprefix\url{https://doi.org/10.1063/1.4953071}.

\bibitem[{\citenamefont{Solak}(2006)}]{Solak_2006}
\bibinfo{author}{\bibfnamefont{H.~H.} \bibnamefont{Solak}},
  \bibinfo{journal}{Journal of Physics D: Applied Physics}
  \textbf{\bibinfo{volume}{39}}, \bibinfo{pages}{R171} (\bibinfo{year}{2006}).

\bibitem[{\citenamefont{McNeil and Thompson}(2010)}]{McNeil2010}
\bibinfo{author}{\bibfnamefont{B.~W.~J.} \bibnamefont{McNeil}}
  \bibnamefont{and} \bibinfo{author}{\bibfnamefont{N.~R.}
  \bibnamefont{Thompson}}, \bibinfo{journal}{Nature Photonics}
  \textbf{\bibinfo{volume}{4}}, \bibinfo{pages}{814} (\bibinfo{year}{2010}),
  \urlprefix\url{https://doi.org/10.1038/nphoton.2010.239}.

\bibitem[{\citenamefont{Pellegrini et~al.}(2016)\citenamefont{Pellegrini,
  Marinelli, and Reiche}}]{RevModPhys.88.015006}
\bibinfo{author}{\bibfnamefont{C.}~\bibnamefont{Pellegrini}},
  \bibinfo{author}{\bibfnamefont{A.}~\bibnamefont{Marinelli}},
  \bibnamefont{and} \bibinfo{author}{\bibfnamefont{S.}~\bibnamefont{Reiche}},
  \bibinfo{journal}{Rev. Mod. Phys.} \textbf{\bibinfo{volume}{88}},
  \bibinfo{pages}{015006} (\bibinfo{year}{2016}),
  \urlprefix\url{https://link.aps.org/doi/10.1103/RevModPhys.88.015006}.

\bibitem[{\citenamefont{Liu et~al.}(2017)\citenamefont{Liu, Xiao, Ye, Wang,
  Cui, Feng, Zhang, and Huang}}]{Liu2017}
\bibinfo{author}{\bibfnamefont{F.}~\bibnamefont{Liu}},
  \bibinfo{author}{\bibfnamefont{L.}~\bibnamefont{Xiao}},
  \bibinfo{author}{\bibfnamefont{Y.}~\bibnamefont{Ye}},
  \bibinfo{author}{\bibfnamefont{M.}~\bibnamefont{Wang}},
  \bibinfo{author}{\bibfnamefont{K.}~\bibnamefont{Cui}},
  \bibinfo{author}{\bibfnamefont{X.}~\bibnamefont{Feng}},
  \bibinfo{author}{\bibfnamefont{W.}~\bibnamefont{Zhang}}, \bibnamefont{and}
  \bibinfo{author}{\bibfnamefont{Y.}~\bibnamefont{Huang}},
  \bibinfo{journal}{Nature Photonics} \textbf{\bibinfo{volume}{11}},
  \bibinfo{pages}{289} (\bibinfo{year}{2017}),
  \urlprefix\url{https://doi.org/10.1038/nphoton.2017.45}.

\bibitem[{\citenamefont{Adamo et~al.}(2009)\citenamefont{Adamo, MacDonald, Fu,
  Wang, Tsai, Garc\'{\i}a~de Abajo, and Zheludev}}]{PhysRevLett.103.113901}
\bibinfo{author}{\bibfnamefont{G.}~\bibnamefont{Adamo}},
  \bibinfo{author}{\bibfnamefont{K.~F.} \bibnamefont{MacDonald}},
  \bibinfo{author}{\bibfnamefont{Y.~H.} \bibnamefont{Fu}},
  \bibinfo{author}{\bibfnamefont{C.-M.} \bibnamefont{Wang}},
  \bibinfo{author}{\bibfnamefont{D.~P.} \bibnamefont{Tsai}},
  \bibinfo{author}{\bibfnamefont{F.~J.} \bibnamefont{Garc\'{\i}a~de Abajo}},
  \bibnamefont{and} \bibinfo{author}{\bibfnamefont{N.~I.}
  \bibnamefont{Zheludev}}, \bibinfo{journal}{Phys. Rev. Lett.}
  \textbf{\bibinfo{volume}{103}}, \bibinfo{pages}{113901}
  (\bibinfo{year}{2009}),
  \urlprefix\url{https://link.aps.org/doi/10.1103/PhysRevLett.103.113901}.

\bibitem[{\citenamefont{Massuda et~al.}(2017)\citenamefont{Massuda,
  Roques-Carmes, Solanki, Yang, Kooi, Habbal, Kaminer, and
  Solja\v{c}i\'{c}}}]{Massuda:17}
\bibinfo{author}{\bibfnamefont{A.}~\bibnamefont{Massuda}},
  \bibinfo{author}{\bibfnamefont{C.}~\bibnamefont{Roques-Carmes}},
  \bibinfo{author}{\bibfnamefont{A.}~\bibnamefont{Solanki}},
  \bibinfo{author}{\bibfnamefont{Y.}~\bibnamefont{Yang}},
  \bibinfo{author}{\bibfnamefont{S.~E.} \bibnamefont{Kooi}},
  \bibinfo{author}{\bibfnamefont{F.}~\bibnamefont{Habbal}},
  \bibinfo{author}{\bibfnamefont{I.}~\bibnamefont{Kaminer}}, \bibnamefont{and}
  \bibinfo{author}{\bibfnamefont{M.}~\bibnamefont{Solja\v{c}i\'{c}}}, in
  \emph{\bibinfo{booktitle}{Conference on Lasers and Electro-Optics}}
  (\bibinfo{publisher}{Optical Society of America}, \bibinfo{year}{2017}), p.
  \bibinfo{pages}{JTh5B.8},
  \urlprefix\url{http://www.osapublishing.org/abstract.cfm?URI=CLEO_SI-2017-JTh5B.8}.

\bibitem[{\citenamefont{Massuda et~al.}(2018)\citenamefont{Massuda,
  Roques-Carmes, Yang, Kooi, Yang, Murdia, Berggren, Kaminer, and
  Soljačić}}]{smith_purcell_low_energy}
\bibinfo{author}{\bibfnamefont{A.}~\bibnamefont{Massuda}},
  \bibinfo{author}{\bibfnamefont{C.}~\bibnamefont{Roques-Carmes}},
  \bibinfo{author}{\bibfnamefont{Y.}~\bibnamefont{Yang}},
  \bibinfo{author}{\bibfnamefont{S.~E.} \bibnamefont{Kooi}},
  \bibinfo{author}{\bibfnamefont{Y.}~\bibnamefont{Yang}},
  \bibinfo{author}{\bibfnamefont{C.}~\bibnamefont{Murdia}},
  \bibinfo{author}{\bibfnamefont{K.~K.} \bibnamefont{Berggren}},
  \bibinfo{author}{\bibfnamefont{I.}~\bibnamefont{Kaminer}}, \bibnamefont{and}
  \bibinfo{author}{\bibfnamefont{M.}~\bibnamefont{Soljačić}},
  \bibinfo{journal}{ACS Photonics} \textbf{\bibinfo{volume}{5}},
  \bibinfo{pages}{3513} (\bibinfo{year}{2018}),
  \urlprefix\url{https://doi.org/10.1021/acsphotonics.8b00743}.

\bibitem[{\citenamefont{Kaminer et~al.}(2017)\citenamefont{Kaminer, Kooi,
  Shiloh, Zhen, Shen, L\'opez, Remez, Skirlo, Yang, Joannopoulos
  et~al.}}]{PhysRevX.7.011003}
\bibinfo{author}{\bibfnamefont{I.}~\bibnamefont{Kaminer}},
  \bibinfo{author}{\bibfnamefont{S.~E.} \bibnamefont{Kooi}},
  \bibinfo{author}{\bibfnamefont{R.}~\bibnamefont{Shiloh}},
  \bibinfo{author}{\bibfnamefont{B.}~\bibnamefont{Zhen}},
  \bibinfo{author}{\bibfnamefont{Y.}~\bibnamefont{Shen}},
  \bibinfo{author}{\bibfnamefont{J.~J.} \bibnamefont{L\'opez}},
  \bibinfo{author}{\bibfnamefont{R.}~\bibnamefont{Remez}},
  \bibinfo{author}{\bibfnamefont{S.~A.} \bibnamefont{Skirlo}},
  \bibinfo{author}{\bibfnamefont{Y.}~\bibnamefont{Yang}},
  \bibinfo{author}{\bibfnamefont{J.~D.} \bibnamefont{Joannopoulos}},
  \bibnamefont{et~al.}, \bibinfo{journal}{Phys. Rev. X}
  \textbf{\bibinfo{volume}{7}}, \bibinfo{pages}{011003} (\bibinfo{year}{2017}),
  \urlprefix\url{https://link.aps.org/doi/10.1103/PhysRevX.7.011003}.

\bibitem[{\citenamefont{Roques-Carmes et~al.}(2019)\citenamefont{Roques-Carmes,
  Kooi, Yang, Massuda, Keathley, Zaidi, Yang, Joannopoulos, Berggren, Kaminer
  et~al.}}]{Roques-Carmes2019}
\bibinfo{author}{\bibfnamefont{C.}~\bibnamefont{Roques-Carmes}},
  \bibinfo{author}{\bibfnamefont{S.~E.} \bibnamefont{Kooi}},
  \bibinfo{author}{\bibfnamefont{Y.}~\bibnamefont{Yang}},
  \bibinfo{author}{\bibfnamefont{A.}~\bibnamefont{Massuda}},
  \bibinfo{author}{\bibfnamefont{P.~D.} \bibnamefont{Keathley}},
  \bibinfo{author}{\bibfnamefont{A.}~\bibnamefont{Zaidi}},
  \bibinfo{author}{\bibfnamefont{Y.}~\bibnamefont{Yang}},
  \bibinfo{author}{\bibfnamefont{J.~D.} \bibnamefont{Joannopoulos}},
  \bibinfo{author}{\bibfnamefont{K.~K.} \bibnamefont{Berggren}},
  \bibinfo{author}{\bibfnamefont{I.}~\bibnamefont{Kaminer}},
  \bibnamefont{et~al.}, \bibinfo{journal}{Nature Communications}
  \textbf{\bibinfo{volume}{10}}, \bibinfo{pages}{3176} (\bibinfo{year}{2019}),
  ISSN \bibinfo{issn}{2041-1723},
  \urlprefix\url{https://doi.org/10.1038/s41467-019-11070-7}.

\bibitem[{\citenamefont{Yang et~al.}(2018)\citenamefont{Yang, Massuda,
  Roques-Carmes, Kooi, Christensen, Johnson, Joannopoulos, Miller, Kaminer, and
  Soljacic}}]{Yang2018}
\bibinfo{author}{\bibfnamefont{Y.}~\bibnamefont{Yang}},
  \bibinfo{author}{\bibfnamefont{A.}~\bibnamefont{Massuda}},
  \bibinfo{author}{\bibfnamefont{C.}~\bibnamefont{Roques-Carmes}},
  \bibinfo{author}{\bibfnamefont{S.~E.} \bibnamefont{Kooi}},
  \bibinfo{author}{\bibfnamefont{T.}~\bibnamefont{Christensen}},
  \bibinfo{author}{\bibfnamefont{S.~G.} \bibnamefont{Johnson}},
  \bibinfo{author}{\bibfnamefont{J.~D.} \bibnamefont{Joannopoulos}},
  \bibinfo{author}{\bibfnamefont{O.~D.} \bibnamefont{Miller}},
  \bibinfo{author}{\bibfnamefont{I.}~\bibnamefont{Kaminer}}, \bibnamefont{and}
  \bibinfo{author}{\bibfnamefont{M.}~\bibnamefont{Soljacic}},
  \bibinfo{journal}{Nature Physics} \textbf{\bibinfo{volume}{14}},
  \bibinfo{pages}{894} (\bibinfo{year}{2018}), ISSN \bibinfo{issn}{1745-2481},
  \urlprefix\url{https://doi.org/10.1038/s41567-018-0180-2}.

\bibitem[{\citenamefont{Wong et~al.}(2015)\citenamefont{Wong, Kaminer, Ilic,
  Joannopoulos, and Soljacic}}]{Wong2015}
\bibinfo{author}{\bibfnamefont{L.~J.} \bibnamefont{Wong}},
  \bibinfo{author}{\bibfnamefont{I.}~\bibnamefont{Kaminer}},
  \bibinfo{author}{\bibfnamefont{O.}~\bibnamefont{Ilic}},
  \bibinfo{author}{\bibfnamefont{J.~D.} \bibnamefont{Joannopoulos}},
  \bibnamefont{and} \bibinfo{author}{\bibfnamefont{M.}~\bibnamefont{Soljacic}},
  \bibinfo{journal}{Nature Photonics} \textbf{\bibinfo{volume}{10}},
  \bibinfo{pages}{46} (\bibinfo{year}{2015}),
  \urlprefix\url{https://doi.org/10.1038/nphoton.2015.223}.

\bibitem[{\citenamefont{Rosolen et~al.}(2018)\citenamefont{Rosolen, Wong,
  Rivera, Maes, Soljacic, and Kaminer}}]{Rosolen2018}
\bibinfo{author}{\bibfnamefont{G.}~\bibnamefont{Rosolen}},
  \bibinfo{author}{\bibfnamefont{L.~J.} \bibnamefont{Wong}},
  \bibinfo{author}{\bibfnamefont{N.}~\bibnamefont{Rivera}},
  \bibinfo{author}{\bibfnamefont{B.}~\bibnamefont{Maes}},
  \bibinfo{author}{\bibfnamefont{M.}~\bibnamefont{Soljacic}}, \bibnamefont{and}
  \bibinfo{author}{\bibfnamefont{I.}~\bibnamefont{Kaminer}},
  \bibinfo{journal}{Light: Science \& Applications}
  \textbf{\bibinfo{volume}{7}}, \bibinfo{pages}{64} (\bibinfo{year}{2018}),
  ISSN \bibinfo{issn}{2047-7538},
  \urlprefix\url{https://doi.org/10.1038/s41377-018-0065-2}.

\bibitem[{\citenamefont{Salasyuk et~al.}(2018)\citenamefont{Salasyuk,
  Rudkovskaya, Danilov, Glavin, Kukhtaruk, Wang, Rushforth, Nekludova, Sokolov,
  Elistratov et~al.}}]{PhysRevB.97.060404}
\bibinfo{author}{\bibfnamefont{A.~S.} \bibnamefont{Salasyuk}},
  \bibinfo{author}{\bibfnamefont{A.~V.} \bibnamefont{Rudkovskaya}},
  \bibinfo{author}{\bibfnamefont{A.~P.} \bibnamefont{Danilov}},
  \bibinfo{author}{\bibfnamefont{B.~A.} \bibnamefont{Glavin}},
  \bibinfo{author}{\bibfnamefont{S.~M.} \bibnamefont{Kukhtaruk}},
  \bibinfo{author}{\bibfnamefont{M.}~\bibnamefont{Wang}},
  \bibinfo{author}{\bibfnamefont{A.~W.} \bibnamefont{Rushforth}},
  \bibinfo{author}{\bibfnamefont{P.~A.} \bibnamefont{Nekludova}},
  \bibinfo{author}{\bibfnamefont{S.~V.} \bibnamefont{Sokolov}},
  \bibinfo{author}{\bibfnamefont{A.~A.} \bibnamefont{Elistratov}},
  \bibnamefont{et~al.}, \bibinfo{journal}{Phys. Rev. B}
  \textbf{\bibinfo{volume}{97}}, \bibinfo{pages}{060404}
  (\bibinfo{year}{2018}),
  \urlprefix\url{https://link.aps.org/doi/10.1103/PhysRevB.97.060404}.

\bibitem[{\citenamefont{Gubbiotti et~al.}(2005)\citenamefont{Gubbiotti, Tacchi,
  Carlotti, Vavassori, Singh, Goolaup, Adeyeye, Stashkevich, and
  Kostylev}}]{PhysRevB.72.224413}
\bibinfo{author}{\bibfnamefont{G.}~\bibnamefont{Gubbiotti}},
  \bibinfo{author}{\bibfnamefont{S.}~\bibnamefont{Tacchi}},
  \bibinfo{author}{\bibfnamefont{G.}~\bibnamefont{Carlotti}},
  \bibinfo{author}{\bibfnamefont{P.}~\bibnamefont{Vavassori}},
  \bibinfo{author}{\bibfnamefont{N.}~\bibnamefont{Singh}},
  \bibinfo{author}{\bibfnamefont{S.}~\bibnamefont{Goolaup}},
  \bibinfo{author}{\bibfnamefont{A.~O.} \bibnamefont{Adeyeye}},
  \bibinfo{author}{\bibfnamefont{A.}~\bibnamefont{Stashkevich}},
  \bibnamefont{and} \bibinfo{author}{\bibfnamefont{M.}~\bibnamefont{Kostylev}},
  \bibinfo{journal}{Phys. Rev. B} \textbf{\bibinfo{volume}{72}},
  \bibinfo{pages}{224413} (\bibinfo{year}{2005}),
  \urlprefix\url{https://link.aps.org/doi/10.1103/PhysRevB.72.224413}.

\bibitem[{\citenamefont{{Adeyeye} et~al.}(2008)\citenamefont{{Adeyeye},
  {Goolaup}, {Singh}, {Jun}, {Wang}, {Jain}, and {Tripathy}}}]{4544892}
\bibinfo{author}{\bibfnamefont{A.~O.} \bibnamefont{{Adeyeye}}},
  \bibinfo{author}{\bibfnamefont{S.}~\bibnamefont{{Goolaup}}},
  \bibinfo{author}{\bibfnamefont{N.}~\bibnamefont{{Singh}}},
  \bibinfo{author}{\bibfnamefont{W.}~\bibnamefont{{Jun}}},
  \bibinfo{author}{\bibfnamefont{C.~C.} \bibnamefont{{Wang}}},
  \bibinfo{author}{\bibfnamefont{S.}~\bibnamefont{{Jain}}}, \bibnamefont{and}
  \bibinfo{author}{\bibfnamefont{D.}~\bibnamefont{{Tripathy}}},
  \bibinfo{journal}{IEEE Transactions on Magnetics}
  \textbf{\bibinfo{volume}{44}}, \bibinfo{pages}{1935} (\bibinfo{year}{2008}),
  ISSN \bibinfo{issn}{0018-9464}.

\bibitem[{\citenamefont{Buchholz et~al.}(2006)\citenamefont{Buchholz,
  Drouvelis, and Schmelcher}}]{PhysRevB.73.235346}
\bibinfo{author}{\bibfnamefont{D.}~\bibnamefont{Buchholz}},
  \bibinfo{author}{\bibfnamefont{P.~S.} \bibnamefont{Drouvelis}},
  \bibnamefont{and}
  \bibinfo{author}{\bibfnamefont{P.}~\bibnamefont{Schmelcher}},
  \bibinfo{journal}{Phys. Rev. B} \textbf{\bibinfo{volume}{73}},
  \bibinfo{pages}{235346} (\bibinfo{year}{2006}),
  \urlprefix\url{https://link.aps.org/doi/10.1103/PhysRevB.73.235346}.

\bibitem[{\citenamefont{Mart\'{i}n et~al.}(2003)\citenamefont{Mart\'{i}n,
  Nogu\'{e}s, Liu, Vicent, and Schuller}}]{magnetic_nanostructures}
\bibinfo{author}{\bibfnamefont{J.}~\bibnamefont{Mart\'{i}n}},
  \bibinfo{author}{\bibfnamefont{J.}~\bibnamefont{Nogu\'{e}s}},
  \bibinfo{author}{\bibfnamefont{K.}~\bibnamefont{Liu}},
  \bibinfo{author}{\bibfnamefont{J.}~\bibnamefont{Vicent}}, \bibnamefont{and}
  \bibinfo{author}{\bibfnamefont{I.~K.} \bibnamefont{Schuller}},
  \bibinfo{journal}{Journal of Magnetism and Magnetic Materials}
  \textbf{\bibinfo{volume}{256}}, \bibinfo{pages}{449} (\bibinfo{year}{2003}),
  \urlprefix\url{"http://www.sciencedirect.com/science/article/pii/S0304885302008983"}.

\bibitem[{\citenamefont{Luttge}(2009)}]{Luttge_2009}
\bibinfo{author}{\bibfnamefont{R.}~\bibnamefont{Luttge}},
  \bibinfo{journal}{Journal of Physics D: Applied Physics}
  \textbf{\bibinfo{volume}{42}}, \bibinfo{pages}{123001}
  (\bibinfo{year}{2009}).

\bibitem[{\citenamefont{Manfrinato et~al.}(2017)\citenamefont{Manfrinato,
  Stein, Zhang, Nam, Yager, Stach, and Black}}]{Manfrinato2017}
\bibinfo{author}{\bibfnamefont{V.~R.} \bibnamefont{Manfrinato}},
  \bibinfo{author}{\bibfnamefont{A.}~\bibnamefont{Stein}},
  \bibinfo{author}{\bibfnamefont{L.}~\bibnamefont{Zhang}},
  \bibinfo{author}{\bibfnamefont{C.-Y.} \bibnamefont{Nam}},
  \bibinfo{author}{\bibfnamefont{K.~G.} \bibnamefont{Yager}},
  \bibinfo{author}{\bibfnamefont{E.~A.} \bibnamefont{Stach}}, \bibnamefont{and}
  \bibinfo{author}{\bibfnamefont{C.~T.} \bibnamefont{Black}},
  \bibinfo{journal}{Nano Letters} \textbf{\bibinfo{volume}{17}},
  \bibinfo{pages}{4562} (\bibinfo{year}{2017}), ISSN \bibinfo{issn}{1530-6984},
  \urlprefix\url{https://doi.org/10.1021/acs.nanolett.7b00514}.

\bibitem[{\citenamefont{Wong et~al.}(2017)\citenamefont{Wong, Hong, Carbajo,
  Fallahi, Piot, Soljacic, Joannopoulos, K{\"a}rtner, and Kaminer}}]{Wong2017}
\bibinfo{author}{\bibfnamefont{L.~J.} \bibnamefont{Wong}},
  \bibinfo{author}{\bibfnamefont{K.-H.} \bibnamefont{Hong}},
  \bibinfo{author}{\bibfnamefont{S.}~\bibnamefont{Carbajo}},
  \bibinfo{author}{\bibfnamefont{A.}~\bibnamefont{Fallahi}},
  \bibinfo{author}{\bibfnamefont{P.}~\bibnamefont{Piot}},
  \bibinfo{author}{\bibfnamefont{M.}~\bibnamefont{Soljacic}},
  \bibinfo{author}{\bibfnamefont{J.~D.} \bibnamefont{Joannopoulos}},
  \bibinfo{author}{\bibfnamefont{F.~X.} \bibnamefont{K{\"a}rtner}},
  \bibnamefont{and} \bibinfo{author}{\bibfnamefont{I.}~\bibnamefont{Kaminer}},
  \bibinfo{journal}{Scientific Reports} \textbf{\bibinfo{volume}{7}},
  \bibinfo{pages}{11159} (\bibinfo{year}{2017}), ISSN
  \bibinfo{issn}{2045-2322},
  \urlprefix\url{https://doi.org/10.1038/s41598-017-11547-9}.

\bibitem[{\citenamefont{Bogner et~al.}(2007)\citenamefont{Bogner, Jouneau,
  Thollet, Basset, and Gauthier}}]{wet-STEM}
\bibinfo{author}{\bibfnamefont{A.}~\bibnamefont{Bogner}},
  \bibinfo{author}{\bibfnamefont{P.-H.} \bibnamefont{Jouneau}},
  \bibinfo{author}{\bibfnamefont{G.}~\bibnamefont{Thollet}},
  \bibinfo{author}{\bibfnamefont{D.}~\bibnamefont{Basset}}, \bibnamefont{and}
  \bibinfo{author}{\bibfnamefont{C.}~\bibnamefont{Gauthier}},
  \bibinfo{journal}{Micron} \textbf{\bibinfo{volume}{38}}, \bibinfo{pages}{390}
  (\bibinfo{year}{2007}),
  \urlprefix\url{https://doi.org/10.1016/j.micron.2006.06.008}.

\bibitem[{\citenamefont{Goldstein et~al.}(2007)\citenamefont{Goldstein,
  Newbury, Joy, Lyman, Echlin, Lifshin, Sawyer, and Michael}}]{xray}
\bibinfo{author}{\bibfnamefont{J.}~\bibnamefont{Goldstein}},
  \bibinfo{author}{\bibfnamefont{D.~E.} \bibnamefont{Newbury}},
  \bibinfo{author}{\bibfnamefont{D.~C.} \bibnamefont{Joy}},
  \bibinfo{author}{\bibfnamefont{C.~E.} \bibnamefont{Lyman}},
  \bibinfo{author}{\bibfnamefont{P.}~\bibnamefont{Echlin}},
  \bibinfo{author}{\bibfnamefont{E.}~\bibnamefont{Lifshin}},
  \bibinfo{author}{\bibfnamefont{L.}~\bibnamefont{Sawyer}}, \bibnamefont{and}
  \bibinfo{author}{\bibfnamefont{J.}~\bibnamefont{Michael}},
  \emph{\bibinfo{title}{Scanning Electron Microscopy and X-Ray Microanalysis}}
  (\bibinfo{publisher}{Springer}, \bibinfo{year}{2007}), \bibinfo{edition}{3rd}
  ed., ISBN \bibinfo{isbn}{0306472929}.

\bibitem[{\citenamefont{Brinkmann et~al.}(2001)\citenamefont{Brinkmann,
  Derbenev, and Fl\"ottmann}}]{PhysRevSTAB.4.053501}
\bibinfo{author}{\bibfnamefont{R.}~\bibnamefont{Brinkmann}},
  \bibinfo{author}{\bibfnamefont{Y.}~\bibnamefont{Derbenev}}, \bibnamefont{and}
  \bibinfo{author}{\bibfnamefont{K.}~\bibnamefont{Fl\"ottmann}},
  \bibinfo{journal}{Phys. Rev. ST Accel. Beams} \textbf{\bibinfo{volume}{4}},
  \bibinfo{pages}{053501} (\bibinfo{year}{2001}),
  \urlprefix\url{https://link.aps.org/doi/10.1103/PhysRevSTAB.4.053501}.

\bibitem[{\citenamefont{Piot et~al.}(2006)\citenamefont{Piot, Sun, and
  Kim}}]{PhysRevSTAB.9.031001}
\bibinfo{author}{\bibfnamefont{P.}~\bibnamefont{Piot}},
  \bibinfo{author}{\bibfnamefont{Y.-E.} \bibnamefont{Sun}}, \bibnamefont{and}
  \bibinfo{author}{\bibfnamefont{K.-J.} \bibnamefont{Kim}},
  \bibinfo{journal}{Phys. Rev. ST Accel. Beams} \textbf{\bibinfo{volume}{9}},
  \bibinfo{pages}{031001} (\bibinfo{year}{2006}),
  \urlprefix\url{https://link.aps.org/doi/10.1103/PhysRevSTAB.9.031001}.

\bibitem[{\citenamefont{Sears et~al.}(2008)\citenamefont{Sears, Colby,
  Ischebeck, McGuinness, Nelson, Noble, Siemann, Spencer, Walz, Plettner
  et~al.}}]{PhysRevSTAB.11.061301}
\bibinfo{author}{\bibfnamefont{C.~M.~S.} \bibnamefont{Sears}},
  \bibinfo{author}{\bibfnamefont{E.}~\bibnamefont{Colby}},
  \bibinfo{author}{\bibfnamefont{R.}~\bibnamefont{Ischebeck}},
  \bibinfo{author}{\bibfnamefont{C.}~\bibnamefont{McGuinness}},
  \bibinfo{author}{\bibfnamefont{J.}~\bibnamefont{Nelson}},
  \bibinfo{author}{\bibfnamefont{R.}~\bibnamefont{Noble}},
  \bibinfo{author}{\bibfnamefont{R.~H.} \bibnamefont{Siemann}},
  \bibinfo{author}{\bibfnamefont{J.}~\bibnamefont{Spencer}},
  \bibinfo{author}{\bibfnamefont{D.}~\bibnamefont{Walz}},
  \bibinfo{author}{\bibfnamefont{T.}~\bibnamefont{Plettner}},
  \bibnamefont{et~al.}, \bibinfo{journal}{Phys. Rev. ST Accel. Beams}
  \textbf{\bibinfo{volume}{11}}, \bibinfo{pages}{061301}
  (\bibinfo{year}{2008}),
  \urlprefix\url{https://link.aps.org/doi/10.1103/PhysRevSTAB.11.061301}.

\bibitem[{\citenamefont{England et~al.}(2014)\citenamefont{England, Noble,
  Bane, Dowell, Ng, Spencer, Tantawi, Wu, Byer, Peralta
  et~al.}}]{RevModPhys.86.1337}
\bibinfo{author}{\bibfnamefont{R.~J.} \bibnamefont{England}},
  \bibinfo{author}{\bibfnamefont{R.~J.} \bibnamefont{Noble}},
  \bibinfo{author}{\bibfnamefont{K.}~\bibnamefont{Bane}},
  \bibinfo{author}{\bibfnamefont{D.~H.} \bibnamefont{Dowell}},
  \bibinfo{author}{\bibfnamefont{C.-K.} \bibnamefont{Ng}},
  \bibinfo{author}{\bibfnamefont{J.~E.} \bibnamefont{Spencer}},
  \bibinfo{author}{\bibfnamefont{S.}~\bibnamefont{Tantawi}},
  \bibinfo{author}{\bibfnamefont{Z.}~\bibnamefont{Wu}},
  \bibinfo{author}{\bibfnamefont{R.~L.} \bibnamefont{Byer}},
  \bibinfo{author}{\bibfnamefont{E.}~\bibnamefont{Peralta}},
  \bibnamefont{et~al.}, \bibinfo{journal}{Rev. Mod. Phys.}
  \textbf{\bibinfo{volume}{86}}, \bibinfo{pages}{1337} (\bibinfo{year}{2014}),
  \urlprefix\url{https://link.aps.org/doi/10.1103/RevModPhys.86.1337}.

\bibitem[{\citenamefont{Peralta et~al.}(2013)\citenamefont{Peralta, Soong,
  England, Colby, Wu, Montazeri, McGuinness, McNeur, Leedle, Walz
  et~al.}}]{Peralta2013}
\bibinfo{author}{\bibfnamefont{E.~A.} \bibnamefont{Peralta}},
  \bibinfo{author}{\bibfnamefont{K.}~\bibnamefont{Soong}},
  \bibinfo{author}{\bibfnamefont{R.~J.} \bibnamefont{England}},
  \bibinfo{author}{\bibfnamefont{E.~R.} \bibnamefont{Colby}},
  \bibinfo{author}{\bibfnamefont{Z.}~\bibnamefont{Wu}},
  \bibinfo{author}{\bibfnamefont{B.}~\bibnamefont{Montazeri}},
  \bibinfo{author}{\bibfnamefont{C.}~\bibnamefont{McGuinness}},
  \bibinfo{author}{\bibfnamefont{J.}~\bibnamefont{McNeur}},
  \bibinfo{author}{\bibfnamefont{K.~J.} \bibnamefont{Leedle}},
  \bibinfo{author}{\bibfnamefont{D.}~\bibnamefont{Walz}}, \bibnamefont{et~al.},
  \bibinfo{journal}{Nature} \textbf{\bibinfo{volume}{503}}, \bibinfo{pages}{91}
  (\bibinfo{year}{2013}), \urlprefix\url{https://doi.org/10.1038/nature12664}.

\end{thebibliography}

\end{document}